\documentclass[runningheads]{llncs}
\pdfoutput=1

\usepackage{named}
\usepackage{multicols}
\usepackage{times}
\usepackage{helvet}
\usepackage{courier}
\usepackage{amsmath}
\usepackage{graphicx}
\DeclareGraphicsExtensions{.pdf,.png,.jpg,.eps}
\usepackage{breakurl}

\def\ar{\leftarrow}
\def\beq{\begin{equation}}
\def\eeq#1{\label{#1}\end{equation}}
\def\ba{\begin{array}}
\def\ea{\end{array}}
\def\i#1{\hbox{\it #1\/}}
\def\mi#1{{\mathit #1}}
\def\no{\i{not}}
\def\ar{\leftarrow}
\def\rar{\rightarrow}
\def\lrar{\leftrightarrow}

\def\sm{\hbox{\rm SM}}
\def\flp{\hbox{\rm FLP}}

\def\tri{\triangle}

\def\dia{\diamond} 
\def\bfxi{\boldsymbol{\xi}}

\def\mu#1{\mathit{\underline{#1}}}

\def\sum{\text{\sc sum}}

\def\bX{{\bf{x}}}
\def\mvis{\!=\!}

\def\false{\hbox{\sc false}}
\def\true{\hbox{\sc true}}

\long\def\nb#1{}

\long\def\BOC#1\EOC{\message{(Commented text )}}
\long\def\BOCC#1\EOCC{\message{(Commented text )}}
\long\def\BOCCC#1\EOCCC{\message{(Commented text )}}
\long\def\optional#1{}

\newtheorem{example1}{Example} 

\titlerunning{Two New Definitions of Stable Models of LPs with Generalized Quantifiers}
\authorrunning{J.~Lee and Y.~Meng}

\begin{document}

\mainmatter

\title{Two New Definitions of Stable Models of Logic Programs
  with Generalized Quantifiers} 

\author{Joohyung Lee and Yunsong Meng}

\institute{
School of Computing, Informatics and Decision Systems Engineering \\
Arizona State University, Tempe, USA
}

\setcounter{page}{131}

\maketitle

\begin{abstract}
We present alternative definitions of the first-order stable model
semantics and its extension to incorporate generalized quantifiers by
referring to the familiar notion of a reduct instead of referring to
the $\sm$ operator in the original definitions. Also, we extend the FLP
stable model semantics to allow generalized quantifiers by referring
to an operator that is similar to the $\sm$ operator. For a reasonable
syntactic class of logic programs, we show that the two stable model
semantics of generalized quantifiers are interchangeable. 
\end{abstract}

\section{Introduction}

Most versions of the stable model semantics involve grounding. For
instance, according to the FLP semantics
from~\cite{fab04,faber11semantics}, assuming that the domain is 
$\{-1,1,2\}$, program  
\beq
\ba {rcl}
  p(2) &\ \ar\ & \no\ \sum\langle x\!:\!p(x) \rangle\!<\!2 \\
  p(-1) &\ar& \sum\langle x\!:\! p(x)\rangle\!>\!-1\\
  p(1) &\ar& p(-1)\ 
\ea
\eeq{ex1}
is identified with its ground instance w.r.t the domain:
\beq
\ba {rcl}
 p(2) &\ \ar\ & \no\ \sum\langle \{-1\!:\!p(-1), 1\!:\!p(1), 2\!:\!p(2)\}\rangle\!<\! 2 \\
 p(-1) &\ar& \sum\langle \{-1\!:\!p(-1), 1\!:\!p(1), 2\!:\!p(2)\}\rangle\!>\! -1 \\
 p(1) &\ar& p(-1)\ .
\ea
\eeq{ex1-ground}
As described in~\cite{fab04}, it is straightforward to extend the
definition of satisfaction to ground aggregate expressions. For
instance, set $\{p(-1),p(1)\}$ does not satisfy the body of the first
rule of \eqref{ex1-ground}, but satisfies the bodies of the other
rules. The FLP reduct of program~\eqref{ex1-ground} relative to
$\{p(-1),p(1)\}$ consists of the last two rules, and $\{p(-1), p(1)\}$
is its minimal model. Indeed,  $\{p(-1), p(1)\}$ is the only FLP
answer set of program~\eqref{ex1-ground}.

On the other hand, according to the semantics from~\cite{fer05},
program~\eqref{ex1-ground} is identified with some complex
propositional formula containing nested implications:
\[ 
\ba l
  \Big(\neg \big((p(2)\!\rar\! p(-1)\!\lor\! p(1))
 \land (p(1)\!\land\! p(2)\!\rar\! p(-1))
 \land (p(-1)\!\land\! p(1)\!\land\! p(2)\!\rar\!\bot)\big)\rar p(2)\Big) \\
\land\ \Big(\big(p(-1)\!\rar\! p(1)\!\lor\! p(2)\big)\rar p(-1)\Big) \\
\land\ \Big(p(-1)\rar p(1)\Big)\ .
\ea
\] 
Under the stable model semantics of propositional formulas
\cite{fer05}, this formula has two answer sets: $\{p(-1),p(1)\}$ and
$\{p(-1),p(1),p(2)\}$. The relationship between the FLP and the Ferraris
semantics was studied in~\cite{lee09,bartholomew11first-order}.

Unlike the FLP semantics, the definition from \cite{fer05} is not
applicable when the domain is infinite because it would require the
representation of an aggregate expression to involve ``infinite''
conjunctions and disjunctions. This limitation was overcome in the
semantics presented in~\cite{lee09,ferr10}, which extends the
first-order stable model semantics from~\cite{fer07a,ferraris11stable}
to incorporate aggregate expressions. 
Recently, it was further extended to formulas involving generalized
quantifiers \cite{lee12stable1}, which provides a unifying framework
of various extensions of the stable model semantics, including
programs with aggregates, programs with abstract constraint
atoms~\cite{mare04}, and programs with nonmonotonic
dl-atoms~\cite{eiter08combining}. 


In this paper, we revisit the first-order stable model semantics and
its extension to incorporate generalized quantifiers. We provide an
alternative, equivalent definition of a stable model by referring to
grounding and reduct instead of the $\sm$ operator. Our work is
inspired by the work
of Truszczynski~[\citeyear{truszczynski12connecting}], who
introduces infinite conjunctions and disjunctions to account for
grounding quantified sentences. 
Our definition of a stable model can be viewed as a reformulation and
a further generalization of his definition to incorporate generalized
quantifiers.
We define grounding in the same way as done in the FLP
semantics, but define a reduct differently so that the semantics
agrees with the one by Ferraris~[\citeyear{fer05}].
As we explain in Section~\ref{sec:reduct}, our reduct of
program~\eqref{ex1-ground} relative to $\{p(-1),p(1)\}$ is 
\beq
\ba {rcl}
 \bot &\ \ar\ & \bot \\
 p(-1) &\ar& \sum\langle\{-1\!:\!p(-1), 1\!:\!p(1), 2\!:\!\bot)\rangle\} \!>\!-1 \\
 p(1) &\ar& p(-1)\  , 
\ea
\eeq{ex1-reduct1}
which is the program obtained from~(\ref{ex1-ground}) by replacing
each maximal subformula that is not satisfied by $\{p(-1),p(1)\}$
with~$\bot$. Set $\{p(-1), p(1)\}$ is an answer set of
program~(\ref{ex1}) as it is a minimal
model of the reduct. Likewise the reduct relative to
$\{p(-1),p(1),p(2)\}$ is
\[
\ba {rcl}
 p(2) &\ \ar\ & \top \\
 p(-1) &\ar& \sum\{\langle -1\!:\!p(-1), 1\!:\!p(1), 2\!:\!p(2)\rangle\} \!>\!-1 \\
 p(1) &\ar& p(-1)
\ea
\] 
and $\{p(-1),p(1),p(2)\}$ is a minimal model of the program. The
semantics is more direct than the one from~\cite{fer05} as it 
does not involve the complex translation into a propositional formula. 

While the FLP semantics in~\cite{fab04} was defined in the context of
logic programs with aggregates, it can be straightforwardly
extended to allow other ``complex atoms.'' Indeed, the FLP reduct is
the basis of the semantics of HEX programs~\cite{eite05}.
In~\cite{fink10alogical}, the FLP reduct was applied to provide a
semantics of nonmonotonic
dl-programs~\cite{eiter08combining}. In~\cite{bartholomew11first-order},
the FLP semantics of logic programs with aggregates was generalized to
the first-order level. That semantics is defined in terms of
the $\flp$ operator, which is similar to the $\sm$ operator. This paper
further extends the definition to allow generalized quantifiers. 

By providing an alternative definition in the way that the other
semantics was defined, this paper provides a useful insight into the
relationship between the first-order stable model semantics and the
FLP stable model semantics for programs with generalized quantifiers. 
While the two semantics behave differently in the general case, we
show that they coincide on some reasonable syntactic class of logic
programs. This implies that an implementation of one of the semantics
can be viewed as an implementation of the other semantics if we limit
attention to that class of logic programs.



The paper is organized as follows. 
Section~\ref{sec:fosm} reviews the first-order stable model
semantics and its equivalent definition in terms of grounding and
reduct, and Section~\ref{sec:smgq} extends that definition to
incorporate generalized quantifiers. 
Section~\ref{sec:flp} provides an alternative definition of the FLP
semantics with generalized quantifiers via a translation
into second-order formulas. Section~\ref{sec:comparison} compares the
FLP semantics and the first-order stable model semantics in the
general context of programs with generalized quantifiers. 

\section{First-Order Stable Model Semantics}\label{sec:fosm}

\subsection{Review of First-Order Stable Model Semantics}
\label{ssec:review-fosm}

This review follows~\cite{ferraris11stable}, a journal version
of~\cite{fer07a}, which distinguishes between intensional and
non-intensional predicates.

A {\em formula} is defined the same as in first-order logic. A {\em
  signature} consists of {\em function constants} and {\em
  predicate constants}. Function constants of arity $0$ are also
called {\em object constants}.
We assume the following set of primitive propositional connectives
and quantifiers:
$$\bot, \top,\ \land,\ \lor,\ \rar,\ \forall,\ \exists\ .$$
$\neg F$ is an abbreviation of $F\rar\bot$, and $F\lrar G$ stands for
$(F\rar G)\land(G\rar F)$.
We distinguish between atoms and atomic formulas as follows: an {\sl
atom} of a signature~$\sigma$ is an $n$-ary predicate constant
followed by a list of~$n$ terms that can be formed from function
constants in $\sigma$ and object variables; {\sl atomic formulas}
of~$\sigma$ are atoms of~$\sigma$, equalities between terms
of~$\sigma$, and the 0-place connectives~$\bot$ and $\top$.

The stable models of $F$ relative to a list of predicates ${\bf p} =
(p_1,\dots,p_n)$ are defined via the {\em stable model operator with
  the intensional predicates ${\bf p}$}, denoted by $\sm[F; {\bf
  p}]$.\footnote{%
The intensional predicates ${\bf p}$ are the predicates that we
``intend to characterize'' by $F$.}
Let ${\bf u}$ be a list of distinct predicate variables
$u_1,\dots,u_n$. 
By ${\bf u}={\bf p}$ we denote the conjunction of the formulas
$\forall {\bf x}(u_i({\bf x})\lrar p_i({\bf x}))$, where ${\bf x}$ is a
list of distinct object variables of the same length as the arity of
$p_i$, for all $i=1,\dots, n$.  By ${\bf u}\leq{\bf p}$ we denote the
conjunction of the formulas
$\forall {\bf x}(u_i({\bf x})\rar p_i({\bf x}))$ for all
$i=1,\dots, n$, and
${\bf u}<{\bf p}$ stands for
$({\bf u}\leq{\bf p})\land\neg({\bf u}={\bf p})$.
For any first-order sentence~$F$, expression $\sm[F;{\bf p}]$ stands
for the second-order sentence
\[
   F \land \neg \exists {\bf u} (({\bf u}<{\bf p}) \land F^*({\bf u})),
\] 
where $F^*({\bf u})$ is defined recursively:
\begin{itemize}
\item  $p_i({\bf t})^* = u_i({\bf t})$ for any list ${\bf t}$ of terms;
\item  $F^* = F$ for any atomic formula~$F$ 
       that does not contain members of~${\bf p}$;
\item  $(F\land G)^* = F^* \land G^*$;
\item  $(F\lor G)^* = F^* \lor G^*$;
\item  $(F\rar G)^* = (F^* \rar G^*)\land (F \rar G)$;
\item  $(\forall xF)^* = \forall xF^*$;
\item  $(\exists xF)^* = \exists xF^*$.
\end{itemize}

A model of a sentence $F$ (in the sense of first-order logic) is
called {\em ${\bf p}$-stable} if it satisfies $\sm[F; {\bf p}]$.

\begin{example1}\label{ex:2}
Let $F$ be sentence $\forall x (\neg p(x)\rar q(x))$, and let $I$ be an
interpretation whose universe is the set of all nonnegative integers
${\bf N}$, and $p^I(n) = \false$, $q^I(n)=\true$ for all $n\in {\bf
  N}$.  Section~2.4 of \cite{ferraris11stable} tells us that $I$
satisfies $\sm[F; pq]$.
\end{example1}

\subsection{Alternative Definition of First-Order Stable Models via
  Reduct}\label{ssec:reduct-fosm}

For any signature $\sigma$ and its interpretation $I$, by $\sigma^{I}$
we mean the signature obtained from~$\sigma$ by adding new object
constants $\xi^\dia$, called {\em object names}, for every element $\xi$
in the universe of $I$. We identify an interpretation $I$ of $\sigma$
with its extension to $\sigma^{I}$ defined by $I(\xi^\dia) = \xi$.

In order to facilitate defining a reduct, we provide a reformulation of
the standard semantics of first-order logic via ``a ground formula w.r.t. an
interpretation.''

\begin{definition}\label{def:gr-formula}
For any interpretation $I$ of a signature $\sigma$, a {\sl ground formula
  w.r.t.~$I$} is defined recursively as follows.

\begin{itemize}
\item  $p(\xi_1^\dia,\dots,\xi_n^\dia)$, where $p$ is a predicate constant
 of $\sigma$ and $\xi_i^\dia$ are object names of $\sigma^I$, is a
 ground formula w.r.t.~$I$;

\item $\top$ and $\bot$ are ground formulas w.r.t.~$I$;

\item  If $F$ and $G$ are ground formulas w.r.t.~$I$, then 
  $F\land G$, $F\lor G$, $F\rar G$ are ground formulas w.r.t.~$I$;

\item  If $S$ is a set of pairs of the form $\xi^\dia\!\!:\!F$ where
  $\xi^\dia$ is an object name in $\sigma^I$ and $F$ is a ground
  formula w.r.t.~$I$, then $\forall (S)$ and $\exists (S)$ are ground
  formulas w.r.t.~$I$.
\end{itemize}
\end{definition}

The following definition describes a process that turns any first-order
sentence into a ground formula w.r.t. an interpretation: 
\begin{definition}\label{def:gr-fo}
Let $F$ be any first-order sentence of a signature $\sigma$, and let
$I$ be an interpretation of $\sigma$ whose universe is $U$. By
$gr_I[F]$ we denote the ground formula w.r.t.~$I$, which is obtained
by the following process: 
\begin{itemize}
\item  $gr_I[p(t_1,\dots,t_n)] = p((t_1^I)^\dia, \dots, (t_n^I)^\dia)$;
\item  $gr_I[t_1=t_2] = 
       \begin{cases} 
        \top & \text{ if $t_1^I=t_2^I$, and}  \\
        \bot & \text{otherwise};
       \end{cases}
       $
\item  $gr_I[\top] = \top$;\ \ \ \  $gr_I[\bot]=\bot$;
\item  $gr_I[F\odot G]= gr_I[F]\odot gr_I[G]\ \ \ \
 (\odot\in\{\land,\lor,\rar\})$;
\item  $gr_I[Qx F(x)] = Q(\{\xi^\dia\!\!:\!gr_I[F(\xi^\dia)] \mid \xi\in U\})$ \
  \ \ \ ($Q\in\{\forall, \exists\}$).
\end{itemize}
\end{definition}

\begin{definition}\label{def:sat-fo}
For any interpretation $I$ and any ground formula $F$ w.r.t.~$I$, the
truth value of $F$ under $I$, denoted by $F^I$, is defined recursively
as follows.
\begin{itemize}
\item  $p(\xi_1^\dia,\dots,\xi_n^\dia)^I = p^I(\xi_1,\dots,\xi_n)$;
\item  $\top^I=\true$;\ \ \ \  $\bot^I=\false$;
\item  $(F\land G)^I=\true$ iff $F^I=\true$ and $G^I=\true$;
\item  $(F\lor G)^I=\true$ iff $F^I=\true$ or $G^I=\true$;
\item  $(F\rar G)^I=\true$ iff $G^I=\true$ whenever $F^I=\true$;

\item  $\forall (S)^I=\true$ iff the set
       $\{\xi \mid \xi^\dia\!\!:\!F(\xi^\dia)\in S \text{ and }
         F(\xi^\dia)^I=\true\}$ is the same as the universe of $I$; 
\item  $\exists (S)^I=\true$ iff the set 
        $\{\xi \mid \xi^\dia\!\!:\!F(\xi^\dia)\in S \text{ and }
                    F(\xi^\dia)^I=\true\}$ is not empty. 
\end{itemize}
We say that $I$ {\em satisfies} $F$, denoted $I\models F$, if $F^I=\true$.
\end{definition}

\medskip
\noindent{\bf Example~\ref{ex:2} continued (I)}.\ \
{
$gr_I[F]$ is $\forall (\{n^\dia\!\!:\!(\neg p(n^\dia)\rar q(n^\dia))\mid n\in {\bf
  N}\})$.
Clearly, $I$ satisfies $gr_I[F]$.
}\medskip

An interpretation~$I$ of a signature~$\sigma$ can be represented as a
pair $\langle I^\mi{func},I^\mi{pred}\rangle$, where~$I^\mi{func}$ is the
restriction of~$I$ to the function constants of $\sigma$, and 
$I^\mi{pred}$ is the set of atoms, 
formed using predicate constants from~$\sigma$ and the object names
from~$\sigma^I$, which are satisfied by~$I$. For example,
interpretation $I$ in Example~\ref{ex:2} can be represented as
$\langle I^\mi{func},\  \{q(n^\dia)\mid n\in {\bf N}\}\rangle$, 
where $I^\mi{func}$ maps each integer to itself.

The following proposition is immediate from the definitions: 

\begin{proposition}\label{fo-sat}
Let $\sigma$ be a signature that contains finitely many predicate
constants, let $\sigma^\mi{pred}$ be the set of predicate constants
in~$\sigma$, let $I=\langle I^\mi{func},I^\mi{pred}\rangle$ be an
interpretation of $\sigma$, and let $F$ be a first-order sentence of
$\sigma$. Then $I\models F$ iff $I^\mi{pred}\models gr_I[F]$.
\end{proposition}

The introduction of the intermediate form of a ground formula
w.r.t. an interpretation helps us define a reduct. 

\begin{definition}\label{def:reduct-fosm}
For any ground formula $F$ w.r.t.~$I$, the {\em reduct} of $F$
relative to $I$, denoted by $F^\mu{I}$, is obtained by replacing each
maximal subformula that is not satisfied by $I$ with~$\bot$. It can
also be defined recursively as follows.

\begin{itemize}
\item  $
      (p(\xi_1^\dia,\dots,\xi_n^\dia))^\mu{I} =
      \begin{cases}
          p(\xi_1^\dia,\dots,\xi_n^\dia) & \text{if $I\models
            p(\xi_1^\dia,\dots,\xi_n^\dia)$}, \\
          \bot & \text{otherwise;}
      \end{cases}
      $
\item  $\top^\mu{I} = \top;\ \ \ \  \bot^\mu{I}=\bot$;


\item  $(F\odot G)^\mu{I} =
      \begin{cases}
          F^\mu{I}\odot G^\mu{I}  & \text{if $I\models F\odot G$}  \
          \ \ \ (\odot\in\{\land,\lor,\rar\}), \\  
          \bot & \text{otherwise;}
      \end{cases}
      $
\item  $Q (S)^\mu{I} =
     \begin{cases}
          Q (\{\xi^\dia\!\!:\!(F(\xi^\dia))^\mu{I} \mid \xi^\dia\!\!:\!F(\xi^\dia)\in S\}) &
               \text{if $I\models Q (S)$}
          \ \ \ \ (Q\in\{\forall,\exists\}), \\
          \bot & \text{otherwise.}
     \end{cases}\\
     $
\end{itemize}
\end{definition}

The following theorem tells us how first-order stable models can be
characterized in terms of grounding and reduct. 

\begin{theorem}\label{prop:ground-fosm}
Let $\sigma$ be a signature that contains finitely many predicate
constants, let $\sigma^\mi{pred}$ be the set of predicate constants
in~$\sigma$, let $I=\langle I^\mi{func},I^\mi{pred}\rangle$ be an
interpretation of $\sigma$, and let $F$ be a first-order sentence of
$\sigma$. $I$ satisfies $\sm[F;\sigma^\mi{pred}]$ iff $I^\mi{pred}$ is a
minimal set of atoms that satisfies $(\i{gr}_{I}[F])^\mu{I}$.
\end{theorem}

\medskip 
\noindent{\bf Example~\ref{ex:2} continued (II)}.\ \
{\sl 
The reduct of $gr_I[F]$ relative to $I$, $(gr_I[F])^\mu{I}$, is \\ 
$\forall (\{n^\dia\!\!:\!(\neg \bot\rar q(n^\dia))\mid n\in {\bf  N}\})$, which is
equivalent to $\forall (\{n^\dia\!\!:\!q(n^\dia) \mid n\in {\bf N}\})$.
Clearly, \hbox{$I^\mi{pred}=\{q(n^\dia)\mid n\in {\bf N}\}$} is a minimal set of
atoms that satisfies $(gr_I[F])^\mu{I}$.
}\medskip

\subsection{Relation to Infinitary Formulas by Truszczynski}

The definitions of grounding and a reduct in the previous section are
inspired by the work of
Truszczynski~[\citeyear{truszczynski12connecting}], where he
introduces infinite conjunctions and disjunctions to account for the
result of grounding $\forall$ and $\exists$ w.r.t. a given
interpretation. Differences between the two approaches are illustrated
in the following example: 

\begin{example1}
Consider the formula $F=\forall x\ p(x)$ and the interpretation $I$
whose universe is the set of all nonnegative integers ${\bf
  N}$. According to~\cite{truszczynski12connecting}, grounding of $F$
w.r.t. $I$ results in the infinitary propositional formula
\[
   \{ p(n^\dia)\mid n\in N \}^{\land}\ . 
\] 
On the other hand, formula $gr_I[F]$ is 
\[
   \forall (\{n^\dia\!\!:\!p(n^\dia)\mid n\in N\}).
\]
\end{example1}

Our definition of a reduct is essentially equivalent to the
one defined in~\cite{truszczynski12connecting}. In the next section, we
extend our definition to incorporate generalized quantifiers.

\section{Stable Models of Formulas with Generalized
  Quantifiers}\label{sec:smgq} 

\subsection{Review: Formulas with Generalized Quantifiers}

We follow the definition of a formula with generalized quantifiers
from~\cite[Section~5]{wes08} (that is to say, with
Lindstr{\"o}m quantifiers \cite{lindstrom66first} without the isomorphism closure
condition).

We assume a set ${\bf Q}$ of symbols for generalized quantifiers. Each
symbol in ${\bf Q}$ is associated with a tuple of nonnegative integers
$\langle n_1,\dots,n_k \rangle$ ($k\ge 0$, and each $n_i$ is~\hbox{$\ge
0$}), called the {\em type}.
A {\em (GQ-)formula (with the set ${\bf Q}$ of generalized quantifiers)} is
defined in a recursive way:

\begin{itemize}
\item  an atomic formula (in the sense of first-order logic) is a GQ-formula;
\item  if $F_1,\dots, F_k$ ($k\ge 0$) are GQ-formulas and $Q$ is a generalized
  quantifier of type $\langle n_1,\dots,n_k\rangle$ in~${\bf Q}$, then
  \beq
       Q[\bX_1]\dots[\bX_k] (F_1(\bX_1),\dots,F_k(\bX_k))
  \eeq{gq-formula}
  is a GQ-formula, where each $\bX_i$ ($1\le i\le k$) is a list of
  distinct object variables whose length is $n_i$.
\end{itemize}

We say that an occurrence of a variable $x$ in a GQ-formula $F$ is
{\em bound} if it belongs to a subformula of $F$ that has the form
$Q[\bX_1]\dots[\bX_k] (F_1(\bX_1),\dots,F_k(\bX_k))$ such that
$x$ is in some~$\bX_i$.
Otherwise the occurrence is {\em free}. We say that $x$ is {\em free}
in~$F$ if $F$ contains a free occurrence of $x$. A {\sl (GQ-)sentence}
is a GQ-formula with no free variables. 

We assume that ${\bf Q}$ contains type $\langle\rangle$
quantifiers $Q_\bot$ and $Q_\top$,
type $\langle 0,0\rangle$ quantifiers $Q_\land,Q_\lor,Q_\rar$, and
type $\langle 1\rangle$ quantifiers $Q_\forall,Q_\exists$. Each of them
corresponds to the standard logical connectives and quantifiers ---
$\bot,\top,\land,\lor,\rar,\forall,\exists$. These generalized
quantifiers will often be written in the familiar form. For example, we
write $F\land G$ in place of $Q_\land[][](F,G)$, and write
$\forall xF(x)$ in place of $Q_\forall[x] (F(x))$.

As in first-order logic, an interpretation $I$ consists of the
universe $U$ and the evaluation of predicate constants and function
constants. 
For each generalized quantifier $Q$ of type $\langle
n_1,\ldots,n_k\rangle$, $Q^U$ is a function from
${\cal P}(U^{n_1})\times\dots\times {\cal P}(U^{n_k})$ to
$\{\true, \false\}$, where ${\cal P}(U^{n_i})$ denotes
the power set of $U^{n_i}$.


\begin{example1}\label{e-1}
Besides the standard connectives and quantifiers, the following are
some examples of generalized quantifiers.

\begin{itemize}
\item type $\langle 1\rangle$ quantifier $ Q_{\le 2} $ such that
$Q_{\le 2}^{U}(R)=\true$ iff $|R|\le
2$; \ \footnote{It is clear from the type of the quantifier that $R$
  is any subset of $U$. We will skip such explanation.}

\item type $\langle 1\rangle $ quantifier $Q_{majority}$ such
  that
  $Q_{majority}^{U}(R)=\true$  iff
  $|R|> |U\setminus R|$;
\nb{Majority}

\item type $\langle 1,1\rangle$ quantifier $Q_{(\sum, <)}$ 
   such that $Q_{(\sum, <)}^{U}(R_1,R_2) = \true$ iff
 \begin{itemize}
 \item  $\sum(R_1)$ is defined,
 \item  $R_2=\{b\}$, where $b$ is an integer, and
 \item  $\sum(R_1)< b$. 
 \end{itemize}
\end{itemize}
\end{example1}

Given a sentence $F$ of $\sigma^I$, $F^I$ is defined recursively as
follows:
\begin{itemize}
\item  $p(t_1,\dots,t_n)^I=p^I(t_1^I,\dots,t_n^I)$,
\item  $(t_1=t_2)^I= (t_1^I=t_2^I)$,
\item  For a generalized quantifier $Q$ of type
   $\langle n_1,\dots,n_k\rangle$,
   $$
   \ba l
       (Q[\bX_1]\dots[\bX_k](F_1(\bX_1),\dots,F_k(\bX_k)))^I
       = Q^{U}((\bX_1\!:\!F_1(\bX_1))^I,\dots,(\bX_k\!:\!F_k(\bX_k))^I),
   \ea
   $$
       where $(\bX_i\!:\!F_i(\bX_i))^I=\{\bfxi\in U^{n_i} \mid
       (F_i(\bfxi^\dia))^I=\true \}$.
\end{itemize}

We assume that, for the standard logical connectives and quantifiers
$Q$, functions $Q^{U}$ have the standard meaning:
 \begin{multicols}{2}
\begin{itemize}
\item  $Q_\forall^{U}(R)=\true$ iff $R=U$;
\item  $Q_\exists^{U}(R)=\true$ iff $R\cap U\ne\emptyset$;
\item  $Q_\land^{U}(R_1,R_2)=\true$ iff
  $R_1=R_2=\{\epsilon\}$;\footnote{%
$\epsilon$ denotes the empty tuple.
For any interpretation $I$, $U^0=\{\epsilon\}$.
For $I$ to satisfy $Q_{\land}[][](F, G)$,
both $(\epsilon\!:\!F)^I$ and $(\epsilon\!:\!G)^I$ have to be $\{\epsilon\}$,
which means that $F^I=G^I=\true$.
}
\item  $Q_\lor^{U}(R_1,R_2) =\true$ iff $R_1=\{\epsilon\}$ or
  $R_2=\{\epsilon\}$;
\item  $Q_\rar^{U}(R_1,R_2) =\true$ iff $R_1$ is $\emptyset$ or
  $R_2$ is $\{\epsilon\}$;
\item  $Q_\bot^{U}() =\false$; 
\item  $Q_\top^{U}() =\true$.
\end{itemize}
\end{multicols}

We say that an interpretation $I$ {\em satisfies} a GQ-sentence $F$, or
is a {\em model} of $F$, and write $I\models F$, if $F^I = \true$. A
GQ-sentence $F$ is {\sl logically valid} if every interpretation
satisfies $F$. A GQ-formula with free variables is said to be {\sl
  logically valid} if its universal closure is logically valid.

\begin{example1}\label{ex:main}
Program~(\ref{ex1}) in the introduction is identified with the following
GQ-formula~$F_1$: 
\[
\ba l
  (\neg Q_{(\sum, <)} [x][y] (p(x),\ y\mvis 2) \rar p(2))\\
 \land ~(Q_{(\sum, >)} [x][y] (p(x),\ y\mvis -1) \rar p(-1))\\
  \land ~( p(-1)\rar p(1)) \ .
\ea
\]
Consider two Herbrand interpretations of the universe $U=\{-1,1,2\}$:
$I_1=\{p(-1),p(1)\}$ and ~$I_2=\{p(-1),p(1), p(2)\}$. We have
$(Q_{(\sum, <)} [x][y] (p(x),\ y=2))^{I_1}=\true$ since
\begin{itemize}
\item $(x:p(x))^{I_1}=\{-1,1\}$ and $(y: y\mvis 2)^{I_1}=\{2\}$;
\item $Q_{(\sum, <)}^U(\{-1,1\}, \{2\})=\true$.
\end{itemize}
Similarly, $(Q_{(\sum, >)} [x][y] (p(x),\ y\mvis
-1))^{I_2}=\true$ 
since
\begin{itemize}
\item $(x:p(x))^{I_2}=\{-1,1,2\}$ and $(y:y\mvis -1)^{I_2}=\{-1\}$;
\item $Q_{(\sum, >)}^U(\{-1,1,2\}, \{-1\})=\true$.
\end{itemize}
Consequently, both $I_1$ and $I_2$ satisfy $F_1$.
\end{example1}

\subsection{Review: $\sm$-Based Definition of Stable Models of GQ-Formulas}\label{ssec:smgq}

For any GQ-formula $F$ and any list of 
predicates ${\bf p}=(p_1,\dots,p_n)$, formula $\sm[F; {\bf p}]$ is
defined as
\[ 
   F\land\neg\exists {\bf u} (({\bf u}<{\bf p}) \land F^*({\bf u})),
\] 
where $F^*({\bf u})$ is defined recursively:
\begin{itemize}
\item  $p_i({\bf t})^* = u_i({\bf t})$ for any list ${\bf t}$ of terms;
\item  $F^* = F$ for any atomic formula~$F$
       that does not contain members of~${\bf p}$;
\item
\[ 
\ba l
  (Q[\bX_1]\dots[\bX_k] (F_1(\bX_1),\dots,F_k(\bX_k)))^* = \\
  \hspace{3em} Q[\bX_1]\dots[\bX_k] (F_1^*(\bX_1),\dots,F_k^*(\bX_k))
  \land\ Q[\bX_1]\dots[\bX_k] (F_1(\bX_1),\dots,F_k(\bX_k)).
\ea
\] 
\end{itemize}

When $F$ is a sentence, the models of $\sm[F; {\bf p}]$ are called the
{\em ${\bf p}$-stable} models of~$F$: they are the models of $F$ that
are ``stable'' on ${\bf p}$.
We often simply write $\sm[F]$ in place of $\sm[F;{\bf p}]$ when
${\bf p}$ is the list of all predicate constants occurring in~$F$, and
call ${\bf p}$-stable models simply stable models.

As explained in~\cite{lee12stable}, this definition of a stable model
is a proper generalization of the first-order stable model semantics. 

\medskip
\noindent{\bf Example~\ref{ex:main} continued (I)}.\ \
{
For GQ-sentence $F_1$ considered earlier, $\sm[F_1]$ is
\beq
   F_1\land\neg\exists u (u<p \land F_1^*(u))\ ,
\eeq{ex1sm}
where $F_1^*(u)$ is equivalent to the conjunction of $F_1$ and
\[
\ba l
  (\neg Q_{(\sum, <)} [x][y] (p(x), y\mvis 2) \rar u(2))\\
 \land ~((Q_{(\sum, >)} [x][y] (u(x), y\mvis -1)\land ~Q_{(\sum, >)} [x][y] (p(x), y\mvis -1)) \rar u(-1))\\
  \land ~( u(-1)\rar u(1)) \ .
\ea
\] 
The equivalence can be explained by Proposition~1
from~\cite{lee12stable1}, which simplifies the transformation for
monotone and antimonotone GQs. 
$I_1$ and $I_2$ considered earlier satisfy (\ref{ex1sm}) and thus are stable models of $F_1$.
}\medskip

\subsection{Reduct-Based Definition of Stable Models of
  GQ-Formulas}\label{sec:reduct} 

The reduct-based definition of stable models presented in
Section~\ref{ssec:reduct-fosm} can be extended to GQ-formulas as
follows. 

Let $I$ be an interpretation of a signature $\sigma$. 
As before, we assume a set ${\bf Q}$ of generalized quantifiers, which
contains all propositional connectives and standard quantifiers.

\begin{definition}\label{def:gr-gq-formula}
A ground GQ-formula w.r.t. $I$ is defined recursively as follows: 
\begin{itemize}
\item  $p(\xi_1^\dia,\dots,\xi_n^\dia)$, where $p$ is a predicate constant
 of $\sigma$ and $\xi_i^\dia$ are object names of $\sigma^I$, is a
 ground GQ-formula w.r.t.~$I$;
\item  for any $Q\in {\bf Q}$ of type $\langle n_1,\ldots,n_k\rangle$,
  if each $S_i$ is a set of pairs of the form $\bfxi^\dia\!\!:\!F$  
  where $\bfxi^\dia$ is a list of object names from~$\sigma^I$ whose length is
  $n_i$ and $F$ is a ground GQ-formula w.r.t.~$I$, then 
  \[
       Q (S_1,\dots,S_k)
  \] 
  is a ground GQ-formula w.r.t. $I$. 
\end{itemize}
\end{definition} 

The following definition of grounding turns any GQ-sentence
into a ground GQ-formula w.r.t. an interpretation: 
\begin{definition}\label{def:gr-gq}
Let $F$ be a GQ-sentence of a signature $\sigma$, and let $I$ be an
interpretation of~$\sigma$. By $gr_I[F]$ we denote the ground
GQ-formula w.r.t.~$I$ that is obtained by the process similar to the
one in Definition~\ref{def:gr-fo} except that the last two clauses
are replaced by the following single clause: 
\begin{itemize}
\item  
  $ 
   gr_I[Q[\bX_1]\dots[\bX_k] (F_1(\bX_1),\dots,F_k(\bX_k))] = Q (S_1,\dots,S_k)
  $
 
  where $S_i=\{\bfxi^\dia\!\!:\!gr_I[F_i(\bfxi^\dia)]\mid \bfxi^\dia \text{ is a list of
    object names from $\sigma^I$ whose length is $n_i$} \}$.
\end{itemize}
\end{definition}

For any interpretation $I$ and  any ground GQ-formula $F$ w.r.t.~$I$, the
satisfaction relation $I\models F$ is defined recursively as follows.

\begin{definition}\label{def:sat-gq}
For any interpretation $I$ and  any ground GQ-formula $F$ w.r.t.~$I$, the
satisfaction relation $I\models F$ is defined similar to
Definition~\ref{def:sat-fo} except that the last five clauses are
replaced by the following single clause:
\begin{itemize}

\item  
     $Q(S_1,\dots,S_k)^I = Q^U(S_1^I,\dots,S_k^I)$
  where $S_i^I=\{\bfxi \mid \bfxi^\dia\!\!:\!F(\bfxi^\dia)\in S_i,\ F(\bfxi^\dia)^I =\true \}$.
\end{itemize}
\end{definition}

\medskip
\noindent{\bf Example~\ref{ex:main} continued (II)}.\ \
{For Herbrand interpretation $I_1=\{p(-1), p(1)\}$,
formula $gr_{I_1}[F_1]$ is  \footnote{For simplicity, we write $-1, 1,
  2$ instead of their object names $(-1)^\dia, 1^\dia, 2^\dia$.}
\beq
\ba l
  (\neg Q_{(\sum,<)}(\{-1\!:\!p(-1), 1\!:\!p(1), 2\!:\!p(2)\}, \{-1\!:\!\bot,1\!:\!\bot,2\!:\!\top\})\rar p(2)) \\
 \land~ (Q_{(\sum,>)}(\{-1\!:\!p(-1), 1\!:\!p(1), 2\!:\!p(2)\},
 \{-1\!:\!\top,1\!:\!\bot,2\!:\!\bot\})\rar p(-1))\\
 \land~ (p(-1)\rar p(1)) \ .
\ea
\eeq{ex1-gr}
$I_1$ satisfies $Q_{(\sum,<)}(\{-1\!:\!p(-1), 1\!:\!p(1), 2\!:\!p(2)\},
\{-1\!:\!\bot,1\!:\!\bot,2\!:\!\top\})$ because 
$I_1\models p(-1)$, $I_1\models p(1)$, $I_1\not\models p(2)$, and 
$$Q_{(\sum,<)}^U(\{-1,1\}, \{2\})=\true.$$
$I_1$ satisfies $Q_{(\sum,>)}(\{-1\!:\!p(-1), 1\!:\!p(1), 2\!:\!p(2)\},
\{-1\!:\!\top,1\!:\!\bot,2\!:\!\bot\})$ because 
$$Q_{(\sum,>)}^U(\{-1,1\}, \{-1\})=\true.$$
Consequently, $I_1$ satisfies \eqref{ex1-gr}.
}
\medskip


\begin{proposition}\label{prop:gq-sat}
Let $\sigma$ be a signature that contains finitely many predicate
constants, let $\sigma^\mi{pred}$ be the set of predicate constants
in~$\sigma$, let $I=\langle I^\mi{func},I^\mi{pred}\rangle$ be an
interpretation of $\sigma$, and let $F$ be a GQ-sentence of
$\sigma$. Then $I\models F$ iff $I^\mi{pred}\models gr_{I}[F]$.
\end{proposition}

\begin{definition}
For any GQ-formula $F$ w.r.t.~$I$, the reduct of $F$ relative to $I$,
denoted by $F^\mu{I}$, is defined in the same way as in 
Definition~\ref{def:reduct-fosm} by replacing the last two clauses
with the following single clause: 
\begin{itemize} 
\item $(Q (S_1,\dots,S_k))^\mu{I} =
      \begin{cases}
           Q (S_1^\mu{I},\dots,S_k^\mu{I}) & \text{if
             $I\models Q (S_1,\dots,S_k)$}, \\
           \bot & \text{otherwise;}
      \end{cases}\\
      $ 
      where $S_i^\mu{I}=\{\bfxi^\dia\!\!:\!(F(\bfxi^\dia))^\mu{I}\mid
      \bfxi^\dia\!\!:\!F(\bfxi^\dia)\in S_i\}$.
\end{itemize}
\end{definition}

\begin{theorem}\label{prop:ground-smgq}
Let $\sigma$ be a signature that contains finitely many predicate
constants, let $\sigma^\mi{pred}$ be the set of predicate constants
in~$\sigma$, let $I=\langle I^\mi{func},I^\mi{pred}\rangle$
be an interpretation of $\sigma$, and let $F$ be a GQ-sentence of
$\sigma$. $I\models \sm[F;\sigma^\mi{pred}]$ iff $I^\mi{pred}$ is a
minimal set of atoms that satisfies $(\i{gr}_{I}[F])^\mu{I}$\ .
\end{theorem}

\medskip
\noindent{\bf Example~\ref{ex:main} continued (III).}\ \
{
Interpretation $I_1$ considered earlier can be identified with the 
tuple $\langle I^\mi{func},\{p(-1),p(1)\}\rangle$ where $I^\mi{func}$
maps every term to itself. The reduct $(gr_{I_1}[F_1])^\mu{I_1}$ is 
\[
\ba l
  (\bot\rar\bot) \\
 \land~ (Q_{(\sum,>)}(\{-1\!:\!p(-1), 1\!:\!p(1), 2\!:\!\bot\},
        \{-1\!:\!\top, 1\!:\!\bot, 2\!:\!\bot\})\rar p(-1))\\
 \land~ (p(-1)\rar p(1)) \ ,
\ea
\]
which is the GQ-formula representation of \eqref{ex1-reduct1}.
We can check that $\{p(-1),p(1)\}$ is a minimal model of the reduct. 
}
\medskip

Extending Theorem~\ref{prop:ground-smgq} to allow an arbitrary list of
intensional predicates, rather than $\sigma^\mi{pred}$, is
straightforward in view of Proposition~1
from~\cite{lee12reformulating}.

\section{FLP Semantics of Programs with Generalized Quantifiers}
\label{sec:flp}

The FLP stable model semantics~\cite{fab04} is an alternative way to
define stable models. It is the basis of HEX programs, an extension of
the stable model semantics with higher-order and external atoms, which
is implemented in system~{\sc dlv-hex}.
The first-order generalization of the FLP stable model semantics for
programs with aggregates was given in~\cite{bartholomew11first-order},
using the $\flp$ operator that is similar to the $\sm$ operator.
In this section we show how it can be extended to allow
generalized quantifiers. 

\subsection{FLP Semantics of Programs with Generalized Quantifiers}
A {\sl (general) rule} is of the form\beq
  H\ar B
\eeq{rule-f}
where $H$ and $B$ are arbitrary GQ-formulas. A {\sl (general) program}
is a finite set of rules.

Let ${\bf p}$ be a list of distinct predicate constants
$p_1,\dots,p_n$, and let ${\bf u}$ be a list of distinct predicate
variables $u_1,\dots,u_n$. For any formula $G$, formula $G({\bf u})$
is obtained from $G$ by replacing all occurrences of predicates from
${\bf p}$ with the corresponding predicate variables from~${\bf u}$.

Let $\Pi$ be a finite program whose rules have the
form~(\ref{rule-f}). The {\em GQ-representation} $\Pi^{GQ}$
of $\Pi$ is the conjunction of the universal closures of $B\rar H$ for
all rules~(\ref{rule-f}) in $\Pi$. By $\flp[\Pi; {\bf p}]$ we denote
the second-order formula
\[
   \Pi^{GQ}\land\neg\exists {\bf u}({\bf u}<{\bf p}\land
         \Pi^\tri({\bf u}))
\] 
where $\Pi^\tri({\bf u})$ is defined as the conjunction of the
universal closures of
\[
  B\land B({\bf u})\!\rar\! H({\bf u})
\] 
for all rules $H\ar B$ in $\Pi$. 

We will often simply write $\flp[\Pi]$ instead of $\flp[\Pi;{\bf p}]$
when ${\bf p}$ is the list of all predicate constants occurring
in~$\Pi$, and call a model of $\flp[\Pi]$ an {\sl FLP-stable} model
of~$\Pi$.

\medskip
\noindent{\bf Example~\ref{ex:main} continued (IV).}\ \
{
For formula $F_1$ considered earlier, $\flp[F_1]$ is
\beq
   F_1\land\neg\exists u (u<p \land F_1^\tri(u))\ ,
\eeq{ex1flp}
where $F_1^\tri(u)$ is
\[
\ba l
  (\neg Q_{(\sum, <)} [x][y] (p(x), y\mvis 2)\land \neg Q_{(\sum, <)} [x][y] (u(x), y\mvis 2) \rar u(2))\\
 \land ~(Q_{(\sum, >)} [x][y] (p(x), y\mvis -1)\land (Q_{(\sum, >)} [x][y] (u(x), y\mvis -1) \rar u(-1))\\
  \land ~( p(-1)\land u(-1)\rar u(1)) \ .
\ea
\] 
$I_1$ considered earlier satisfies (\ref{ex1flp}) but $I_2$ does not. 
}\medskip

\section{Comparing the FLP Semantics and the First-Order Stable Model Semantics}
\label{sec:comparison}

In this section, we show a class of programs with GQs for which the
FLP semantics and the first-order stable model semantics coincide. 

The following definition is from~\cite{lee12stable}.
We say that a generalized quantifier $Q$ is {\em
  monote in the $i$-th argument position} if the following holds for
any universe $U$: 
if $Q^{U}(R_1,\dots,R_k)=\true$ and
$R_i\subseteq R_i' \subseteq U^{n_i}$,
then 
\[ Q^{U}(R_1,\dots,R_{i-1},R_i',R_{i+1},\dots,R_k)=\true.\]

Consider a program $\Pi$ consisting of rules of the form 
\[
  A_1;\dots; A_l \ar\ E_1,\dots,E_m, \no\ E_{m+1},\dots,\no\ E_n
\] 
($l\ge 0$; $n\ge m\ge 0$), where each $A_i$ is an atomic formula and
each $E_i$ is an atomic formula or a GQ-formula~\eqref{gq-formula}
such that all $F_1(\bX_1),\dots,F_k(\bX_k)$ are atomic
formulas. Furthermore we require that, for every
GQ-formula~\eqref{gq-formula} in one of $E_{m+1},\dots, E_n$, $Q$ is
monotone in all its argument positions.

\begin{proposition}\label{cor:flp-sm-hex}
Let $\Pi$ be a program whose syntax is described as above, and let $F$ be
the GQ-representation of $\Pi$. Then $\flp[\Pi; {\bf p}]$ is
equivalent to $\sm[F; {\bf p}]$.
\end{proposition}

\begin{example1} 
Consider the following one-rule program: 
\beq
   p(a)\ar \no\ Q_{\le 0} [x]\ p(x)\ . 
\eeq{ex2}
This program does not belong to the syntactic class of programs stated
in Proposition~\ref{cor:flp-sm-hex} since $Q_{\le 0} [x]\ p(x)$ is
not monotone in $\{1\}$. Indeed, both $\emptyset$ and $\{p(a)\}$
satisfy $\sm[\Pi; p]$, but only $\emptyset$ satisfies $\flp[\Pi; p]$.
\end{example1} 

Conditions under which the FLP semantics coincides with the
first-order stable model semantics has been studied
in~\cite{lee09,bartholomew11first-order} in the context of logic
programs with aggregates. 

\section{Conclusion}



We introduced two definitions of a stable model. One is a
reformulation of the first-order stable model semantics and its
extension to allow generalized quantifiers by referring to grounding
and reduct, and the other is a reformulation of the FLP semantics and
its extension to allow generalized quantifiers by referring to a
translation into second-order logic. 
These new definitions help us understand the relationship between the
FLP semantics and the first-order stable model semantics, and their
extensions. 
For the class of programs where the two semantics coincide, system
{\sc dlv-hex} can be viewed as an implementation of the stable model
semantics of GQ-formulas; A recent extension of system {\sc
  f2lp}~\cite{lee09a} to allow ``complex'' atoms may be considered as
a front-end to {\sc dlv-hex} to implement the generalized FLP
semantics. 

\section*{Acknowledgements}

We are grateful to Vladimir Lifschitz for useful discussions related
to this paper. We are also grateful to Joseph Babb and the anonymous
referees for their useful comments. This work was partially supported
by the National Science Foundation under Grant IIS-0916116 and by the
South Korea IT R\&D program MKE/KIAT 2010-TD-300404-001.

\bibliographystyle{splncs}



\end{document}